\documentclass[prb,twocolumn,showpacs,superscriptaddress,floatfix]{revtex4}
\usepackage{graphicx,amsfonts,amssymb,amsmath}
\usepackage[pagebackref=true,colorlinks,linkcolor=blue,citecolor=red]{hyperref}

\def\be{\begin{equation}}
\def\ee{\end{equation}}
\def\bea{\begin{eqnarray}}
\def\eea{\end{eqnarray}}

 \newcommand{\Zd}{\mathbb{Z}_2}

 \newcommand{\ket}[1]{|#1\rangle}

 \newcommand{\comutator}[2]{[ #1,#2 ]}

\newcommand{\fe}{\mathfrak e} %
\newcommand{\fv}{\mathfrak v} %
\newcommand{\fp}{\mathfrak p} %
\newcommand{\fP}{\mathfrak P} %
\newcommand{\fU}{\mathfrak{U}} %
\newcommand{\fV}{\mathfrak{V}} %
\newcommand{\id}{1\hspace{-.25em}{\rm I}} %

\begin{document}

\title{Robustness of a topological phase:

Topological color code in parallel magnetic field}

\author{Saeed S. Jahromi}
\email{s.jahromi@dena.kntu.ac.ir}
\affiliation{Department of Physics, K.N. Toosi University of Technology, P.O. Box 15875-4416, Tehran, Iran}
\affiliation{Lehrstuhl f\"{u}r Theoretische Physik I, Otto-Hahn-Stra{\ss}e 4, D-44221 Dortmund, Germany}

\author{Mehdi Kargarian}
\email{kargarian@physics.utexas.edu}
\affiliation{Department of Physics, The University of Texas at Austin, Austin, TX 78712, USA}

\author{S Farhad Masoudi}
\email{masoudi@kntu.ac.ir}
\affiliation{Department of Physics, K.N. Toosi University of Technology, P.O. Box 15875-4416, Tehran, Iran}

\author{Kai Phillip Schmidt}
\email{kai.schmidt@tu-dortmund.de}
\affiliation{Lehrstuhl f\"{u}r Theoretische Physik I, Otto-Hahn-Stra{\ss}e 4, D-44221 Dortmund, Germany}

\begin{abstract}
The robustness of the topological color code, which is a class of error correcting quantum codes,
is investigated under the influence of an uniform magnetic field on the honeycomb lattice. 
Our study relies on two high-order series expansions using perturbative continuous unitary 
transformations in the limit of low and high fields, exact diagonalization and a classical approximation. We show that the topological color code in 
a single parallel field is isospectral to the Baxter-Wu model in a transverse field
on the triangular lattice. It is found that the topological phase is stable up to a critical field beyond which it breaks down 
to the polarized phase by a first-order phase transition. 
The results also suggest that the topological color code is more robust than the toric code, in the parallel magnetic field. 
\end{abstract}
\maketitle

\section{Introduction}
\label{intro}

Conventional phases of matter such as solids or liquids are classically described in the framework of Ginzburg-Landau symmetry breaking theory of
phase transitions \cite{landau}. Over 50 years, it was believed that Landau theory is capable of describing any phases of matter by introducing 
an appropriate local order parameter characterizing different behaviors of two phases on either side of the critical point.
Discovery of the fractional quantum Hall effect \cite{tsui}, introduced an exotic phase of matter carrying a new kind of quantum order, {\it topological order}, which
is beyond the scope of the Landau symmetry breaking theory of phase transitions \cite{wen1, laughlin}.
Since then, investigation of systems with topological order such as quantum dimer models\cite{dimer1, dimer2, dimer3,dimer4}, 
quantum spin models\cite{qu_spin1,qu_spin2,qu_spin3,qu_spin4,qu_spin5,qu_spin6,qu_spin7,qu_spin8,qu_spin9} and superconducting states\cite{superconductor1,superconductor2} 
became a very active area of research in condensed matter physics. Later on, remarkable characteristics of the topologically ordered 
system and their robust nature motivated the physicists in the field of quantum information to introduce new models for fault tolerant quantum computation
\cite{Kitaev1, dennis, bombin1}. 

The toric code by Kitaev is the first example of error correcting quantum models making use of the topologically degenerate ground state 
of the system as a robust quantum memory\cite{Kitaev1}. The ground state degeneracy depends on the topology of the system and it can not be
lifted by any local perturbation to the Hamiltonian. Only non-trivial topological errors are capable of lifting the degenerate ground states 
of the system \cite{wen2}. The ground states are stabilized by a group of local operators called plaquettes and stars making them useful as an 
error correcting quantum code. The gapped excitations above the topologically-ordered ground states correspond to a fascinating class of 
emerging particles that are neither fermions nor bosons. These particles are called anyons. Most interestingly, it has been shown that the non-trivial 
braiding of non-Abelian anyons can be used for fault tolerant quantum computation\cite{leinaas, wilczek}.

Topological color codes are a different class of error correcting quantum codes which allow the implementation of the whole Clifford group 
in a fully topological manner in the ground state space. The latter makes the platform suitable for quantum computation without resorting 
to braiding of quasiparticle excitations \cite{bombin1}. In contrast to the toric code which is built on tetravalent lattices, the color code
is implemented on trivalent structures. Moreover, the toric code belongs to the $\Zd$ gauge symmetry group while the 
topological color code is related to the $\Zd\times\Zd$ gauge symmetry group. This distinction is a direct consequence of the different lattice structure
and the way in which plaquette operators are attached to faces on the lattices. It is worth noting that the topological order in the error correcting quantum codes is directly related to the 
topology of underlying manifold and especially to string structure of states made of plaquette operators. Thus, in contrast to the symmetry protected topological orders (SPT)\cite{STO1}, the topological order in color codes is unrelated to the symmetry. In addition, unlike the SPTs which are
known for gapless excitations on boundary and short-range entanglement\cite{STO2,STO3}, the topological quantum codes have gapped excitations and long-range entanglement.

The robustness of the protected topological order of the topological color code is a very important issue when it comes to practice.
Although the topological degeneracy of the ground state can not be lifted by local perturbations, the question arises as to how much this 
topological nature is robust against local errors and how large a perturbation can be, until the model undergoes a phase transition? 

This question has been extensively studied recently for the case of the conventional toric code in a uniform magnetic 
field \cite{trebst07,vidal09,vidal09b,tupitsyn10,dusuel11,wu12}. It has been shown that the overall phase diagram is very rich 
containing first- and second-order phase transitions, multicriticality \cite{vidal09,tupitsyn10,dusuel11,wu12}, and self-duality \cite{vidal09b} 
depending on the field direction. If the transition is second order, it is typically in the three-dimensional (3D) Ising universality class except on a 
special line in parameter space where a more complicated behavior is detected \cite{dusuel11}. Similarly, the phase diagram 
of a generalized  $\mathbb{Z}_3$ toric code in a field also displays various interesting aspects such as a 3D $XY$ quantum phase 
transition for the simplest case of a single parallel field. \cite{schulz12}.

Such questions have not been studied for the topological color code except a few studies that investigated the 
finite temperature properties of color code\cite{kargarian_finite_temp, helmut_error_tershold}, where it was shown that 
the topological order is stable to some finite temperature depending on size of the system with high error correction threshold. 
It is therefore the main motivation of this 
work to investigate the phase transition of the topological color code in the presence of a magnetic field. This is done 
for a single magnetic field in either the $x$- or $z$-direction on the honeycomb lattice. It is shown that the topological color 
code in a single parallel field is mapped exactly to the Baxter-Wu Hamiltonian \cite{baxter73} in a transverse field on 
the triangular lattice. To the best of our knowledge neither of the models are exactly solvable and the phase diagrams are unknown. 
So already the simplest case of a single parallel field is fundamentally different to the analogous problem for the toric code. 
In a single parallel field, one finds for the latter case a mapping to the conventional transverse field Ising model displaying a 
second-order quantum phase transition\cite{trebst07}.

Here we investigate the topological color code in a single parallel field by high-order series expansion techniques about 
the low- and high-field limit. Additionally, an exact diagonalization as well as a classical approximation for
the Baxter-Wu model in a transverse field were performed. 
Our results clearly reveal that the topological phase breaks down to the polarized phase through a first-order phase transition. 
The behavior of the perturbed topological color code is therefore different to the perturbed toric code. Our results suggest 
that the topological order of the topological color code is more robust.

The outline of the paper is as follows: In Sect.~\ref{TCC}, the topological color codes and their most important properties 
are introduced. In Sec.~\ref{tcc_in_field} a magnetic field is added to the topological color code. 
For the case of a single parallel field, the mapping to the Baxter-Wu model in a transverse field is explicitly given.
We start our discussion on the solution of the problem by exactly diagonalizing the Baxter-Wu Hamiltonian in Sec.~\ref{ed}.
Afterwards in Sect.~\ref{ca}, we present the classical solution of the Baxter-Wu model in a transverse field already 
displaying a first-order phase transition. The high-order series expansions of the low-energy spectrum for both low- and high-field 
limits are the contents of Sect.~\ref{pcut}. The phase diagram of the topological code in a magnetic field is discussed 
in Sect.~\ref{phase_transition}. Finally, Sect.~\ref{conclud} is devoted to the conclusion.
 
\section{Topological Color Code}
\label{TCC}

\begin{figure}
\centerline{\includegraphics[width=\columnwidth]{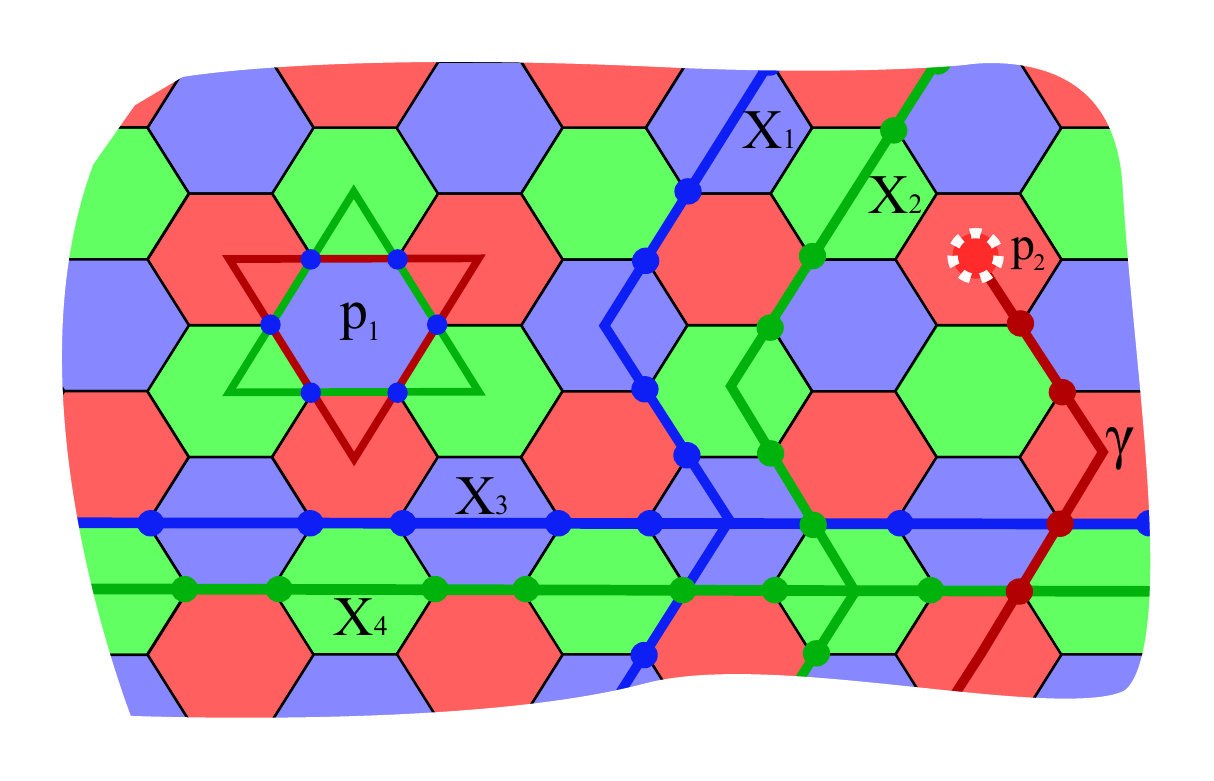}}
\caption{(Color online) A piece of 2-colex on a honeycomb lattice placed on a torus.
Colored circles at the vertices of the lattice represent spins 1/2.
A colored plaquette like $p_1$ is characterized by two closed strings of the other two colors.
$X_1,\ldots,X_4$ are the global strings of the model on a periodic structure such as a torus with $g=1$.
The elementary excitations are described by open strings like $\gamma$ which create particles at their end points on the surface of a 
plaquette (see $p_2$) which share the same color with the string.}
\label{tcctotal}
\end{figure}

Topological color code is a topologically-ordered quantum stabilizer error
correcting code which is defined on a certain class of two-dimensional lattices called 2-colexes \cite{bombin1}.
Local degrees of freedom are spins 1/2 which are placed on the vertices
of an arbitrary 2D trivalent lattice which has 3-colorable faces and can be wrapped around a 
compact surface of arbitrary topology of genus $g$.
Each vertex is connected to three links with different colors say (red, blue, green) and each 
link connects two faces (plaquettes) of the same color.
This colorable structure is denoted by topological color code (TCC) \{$\fv, \fe, \fp$\} where $\fv, \fe$ and $\fp$ correspond 
to a set of vertices, edges and plaquettes respectively. 
An example of a 2-colex construction on a honeycomb lattice is illustrated in Fig.~\ref{tcctotal}.

The protected subspace ${\cal C}$ of the model is determined by the definition of the stabilizer group $\fU$ 
which is generated by a set of plaquette operators. 
For each plaquette $\fp$, there are a pair of stabilizer operators 
$X_{\fp}=\bigotimes_{\fv \in \fp}\sigma_{\fv}^{x}$ and $Z_{\fp}=\bigotimes_{\fv \in \fp} \sigma_{\fv}^{z}$
where $\sigma_{\fv}^{\alpha}$'s are the ordinary Pauli operators. Each plaquette operator satisfies the square identity relation, 
$(X_{\fp})^2 = \id = (Z_{\fp})^2, \forall X_{\fp}, Z_{\fp} \in \fU$, and they mutually commute with each other. The states in protected subspace $\ket{\Psi_{\rm c}}\in {\cal C}$  are
left invariant under the action of the plaquette operators, i.e. 
$X_{\fp} \ket{\Psi_{\rm c}} = \ket{\Psi_{\rm c}},\ \ Z_{\fp} \ket{\Psi_{\rm c}} = \ket{\Psi_{\rm c}}, \ \ \forall \fp \in \fP$.
An erroneous state $\ket{\Psi_{\rm e}}$ is then defined as a state with at least one eigenvalue $-1$ 
which violates the stabilizing condition $X_{\fp} \ket{\Psi_{\rm e}} = - \ket{\Psi_{\rm e}}$
or $Z_{\fp} \ket{\Psi_{\rm e}} = - \ket{\Psi_{\rm e}}$ \cite{bombin2}.

It is possible to construct a Hamiltonian in terms of stabilizer operators with 
topologically degenerate ground states where all erroneous states correspond 
to excited states. Considering a trivalent lattice $\Lambda$, 
the topological color code Hamiltonian reads as:
\be \label{pure_tcc}
H_{\rm TCC}=-J\sum_{\fp \in \Lambda} (X_{\fp} + Z_{\fp}) \quad ,
\ee
where in the following we set $J>0$ to be positive and we focus on the honeycomb lattice. 

\subsection{String-net operators}

Strings are the generalization of plaquette operators which can be either open or closed
and defined by the product of the Pauli spins of that string.
For a generic plaquette, such as a blue one ($\fp_1$) in Fig.~\ref{tcctotal}, two closed strings of
different colors (red and green) which turn around it's boundaries, 
identify the plaquette. In a similar manner, the product of different neighboring
plaquette operators may produce a collection of different boundary operators.
Since all closed strings share an even number of vertices, 
they commute with each other and with the plaquette operators.

There is also another type of string operator which commutes with plaquette 
operators but is not the product of boundary operators.
These operators which are known as the global loops (fundamental cycles), 
depend on the topology of the manifold on which the system is defined and have the 
character of color. 
The number of these closed loops depends on the genus of
the manifold. For example, a torus with $g=1$, 
has two global loops which in contrast to the closed boundaries are non-contractible.

For every homology class of the torus there are two closed strings each of one color 
say green and blue which connect the plaquettes of the same color. 
One can define nontrivial closed strings as \cite{kargarian}: 

\be \label{closed_strings}
{\cal S}_{\mu}^{CX}= \bigotimes_{\fv \in I}\sigma_{\fv}^{x}, \ \ {\cal S}_{\mu}^{CZ}= \bigotimes_{\fv \in I}\sigma_{\fv}^{z} \quad ,
\ee

where $I$ is the set of spins on the string, $\mu=1, 2$ stands for the homology class of the 
torus and $C$ characterizes the color. Following this notation, one can label the non-contractible loops of
the torus as follows:

\be \label{noncontractible_loops}
\begin{gathered}
X_{1} \leftrightarrow {\cal S}_{2}^{BX}, \ \ X_{2} \leftrightarrow {\cal S}_{1}^{GX},\ \ X_{3} \leftrightarrow {\cal S}_{2}^{BX},\ \ X_{4} \leftrightarrow {\cal S}_{1}^{GX}, \\
Z_{1} \leftrightarrow {\cal S}_{1}^{GZ}, \ \ Z_{2} \leftrightarrow {\cal S}_{2}^{BZ},\ \ Z_{3} \leftrightarrow {\cal S}_{1}^{GZ},\ \ Z_{4} \leftrightarrow {\cal S}_{2}^{BZ}.
\end{gathered}
\ee
For the topological color code, these loops are labeled as ($X_1\ldots X_4$) and ($Z_1\ldots Z_4$) and 
shown in Fig.~\ref{tcctotal}. In the next subsection, 
we will show how these operators manifest a 16-dimensional subspace for the protected coding space ${\cal C}$.

\subsection{Ground-state structure}

Since the plaquette operators commute with each other and therefore with the full Hamiltonian \ref{pure_tcc}, they share a simultaneous 
eigenstate and the ground state of Eq.~\ref{pure_tcc} is readily found by finding the eigenstates of the plaquette operators. By starting from a polarized state with all spins in the up direction $\ket{0}^{\otimes |\fV|}$ ($|\fV|$ is the number of vertices on the lattice), 
the ground state of the system
which is stabilized by the plaquette operators is constructed. 
It is convenient to denote the un-normalized ground state in the following form:
\be \label{tcc_ground_state}
\ket{\Psi_{\rm{gs}}}= \prod_{\fp} (1+X_{\fp}) \ket{0}^{\otimes |\fV|}
\ee
This state vector is a superposition of all possible closed string-nets, 
a typical feature of the ground states of systems with topological order \cite{Kitaev1, wen3}.
 
Applying the global loops to the ground state $\ket{\Psi_{\rm{gs}}}$ yields a set of 
different states which satisfy the stabilizing condition. Generically, the set of 
degenerate protected subspace of TCC is defined as:
\be \label{degenerate_gs}
{\cal C}=\{\ket{ijkl}: \ket{ijkl}=X_{1}^{i} X_{2}^{j} X_{3}^{k} X_{4}^{l} \ket{\Psi_{gs}} \}
\ee 
where $X_{1}^{i} X_{2}^{j} X_{3}^{k} X_{4}^{l}$ are defined in Eq.~\ref{noncontractible_loops} and $i,j,k,l=0(1)$ 
correspond to the global string $X$ appearing (not appearing) in the set.
These topological global loops can take one ground state to another one, and any error of this type is
not detectable. A generic state of the ground-state subspace is then a superposition of the different states of the coding space as follows: 
\be \label{degenerate_gs2}
\ket{\Psi} =\sum_{i,j,k,l} a_{i,j,k,l} \ket{ijkl},\ \ \sum_{i,j,k,l} |a_{i,j,k,l}|^{2}=1 \quad.
\ee 
\subsection{Excitations}

Any state $\ket{\Psi}$ of the system which violates the stabilizing condition ($X_{\fp}\ket{\Psi}=\ket{\Psi}$ and
$Z_{\fp}\ket{\Psi}=\ket{\Psi}$), is an excited state. 
For example one can say a state $\ket{\Psi}$ for which $X_{\fp}\ket{\Psi}=-\ket{\Psi}$ ($Z_{\fp}\ket{\Psi}=-\ket{\Psi}$), has an $X$-type ($Z$-type) 
quasiparticle excitation.
Since all of the closed strings commute with each other, it is not
possible to express the excited states of the system in terms of the closed strings.
The elementary excitations of the TCC are described by open strings. Colored strings are open if they have endpoints. 
These endpoints are localized at plaquettes which have the same color as the string, see also Fig.~\ref{tcctotal}. In particular, a plaquette $\fp_2$ is an endpoint of string $\gamma$ if the number of edges of 
$\gamma$ meeting at $\fp_2$ is odd.
In terms of string operators, a plaquette $\fp$ is an endpoint of $\gamma$ if $\{ X_{\gamma} , Z_{\fp}\} = 0$ or, 
equivalently, if $\{ Z_{\gamma} , X_{\fp}\}=0$. Thus, open string operators do not commute with those plaquette 
operators at their ends. As a consequence, the presence of open strings in the system would violate
the stabilizing condition producing an elementary excitation:
\be \label{tcc_excitations}
\begin{gathered}
\ket{\Psi_{\rm{ex}}} =X_{\gamma} \ket{\Psi_{\rm{gs}}} \\ 
Z_{\fp} \ket{\Psi_{\rm{ex}}}= Z_{\fp} X_{\gamma} \ket{\Psi_{\rm{gs}}} =-X_{\gamma} Z_{\fp} \ket{\Psi_{\rm{gs}}}= -\ket{\Psi_{\rm{ex}}}\quad .
\end{gathered}
\ee

These elementary excitations are Abelian anyons. There exist other families of nontrivial topological charges which emerge
by combining the elementary excitations. Each single excitation is a boson by itself, as well as the combination of
two elementary excitations with the same color. Furthermore, combination of the excitations of different color and type creates 
particles with semionic mutual statistics and two families of fermions \cite{bombin3}.
 
Ground states are states where all eigenvalues of plaquette operators are $+1$ for $J>0$. The ground-state energy of the system 
on a lattice with $N_{\fp}$ plaquettes (corresponding to $2N$ sites) is therefore $E_0=-N_{\fp}J=-2NJ$. The energy of excited states depends
 only on the total number of $-1$ eigenvalues of the plaquette operators such that each $-1$ eigenvalue costs the energy $2J$, i.e.~the first 
excited state of the system is $N_{\fp}$-fold degenerate (omitting the topological degeneracy) and it has a total energy $E_1=E_0+2J$. Consequently,
 the energy spectrum is equidistant and the elementary excitations corresponding to $-1$ eigenvalues are static and non-interacting.

\section{Topological Color Code in magnetic Field}
\label{tcc_in_field}
Our major objective is to study the robustness of the TCC in an external magnetic field. We therefore add to the topological color code $H_{\rm TCC}$ a single parallel field in the $x$-direction. Let us stress that it would be fully equivalent to consider the case of a magnetic field in the $z$-direction, because only the behavior of $X$-type and $Z$-type particles are interchanged. To be specific, we study the following Hamiltonian  
\be \label{transverse_tcc}
H=-J\sum_{\fp} (X_{\fp} + Z_{\fp}) -h_x \sum_{\fv} \sigma_{\fv}^x \quad ,
\ee
where the first sum runs over all hexagons and the second one over all sites of the honeycomb lattice. In the following we set $h_x>0$. The plaquette operators read as 
\be \label{plaq_op}
\begin{gathered}
X_{\fp}=\sigma_{1}^{x}\sigma_{2}^{x}\sigma_{3}^{x}\sigma_{4}^{x}\sigma_{5}^{x}\sigma_{6}^{x} \\
Z_{\fp}=\sigma_{1}^{z}\sigma_{2}^{z}\sigma_{3}^{z}\sigma_{4}^{z}\sigma_{5}^{z}\sigma_{6}^{z}
\end{gathered}
\ee
where the numbers ($1, \ldots,6$) represent the six sites of the hexagon $\fp$.

\begin{figure}
\centerline{\includegraphics[width=4cm]{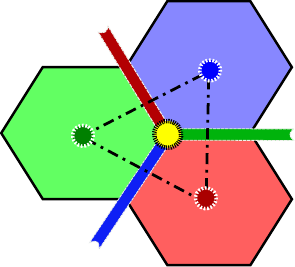}}
\caption{(Color online) Every time a $\sigma_{\fv}^{x}$ acts on the site $\fv$ of the lattice, it can be considered as  
the end point of three different colored open strings say (red, green, blue)
which create three $Z$-type quasiparticle (QP) on the neighboring plaquettes sharing 
the same color with the strings. The yellow circle denotes the $\sigma^{x}$ field 
which is connected to the end points of the red, blue and green strings and the
three colored circles on the surface of the plaquettes are the QP excitations in the topological phase. 
The QPs are created on the vertices of the effective triangular lattice.}
\label{field-pertubation} 
\end{figure}

In the presence of the $x$ field, any trivial closed string ${\cal S}^{cz}=\otimes_{\fv \in s}\sigma_{\fv}^{z}$ pays an excitation energy $2N_{\fv}h_x$ where $N_{\fv}$ is the number
of vertices of the string. The strings stay static and the field perturbation in the Hamiltonian \ref{transverse_tcc} can be considered as a tension term. 
In contrast, the $Z_{\fp}$ plaquette operators deform the strings inducing them to fluctuate. Hence one can think of them as the kinetic term \cite{hamma}. 
As already mentioned in the previous section, the excitations are created by open strings. Every open strings create two excitations at their ends costing a total 
energy of $4J$. In the limit of $J\gg h_x\sim0$, open strings are not energetically favorable and tension is too weak to prevent the closed strings from fluctuating. So as a
result, the ground state of the system is a superposition of strongly fluctuating closed string.
In other limit, $h_x\gg J\sim0$, the strong tension suppresses all of the fluctuating strings and the ground state of the system 
lies in a spin polarized phase where spins point in field direction.
Henceforth, it is physically reasonable to think of a phase transition between the two phases at some critical value of $h_x/J$. 

In order to disclose this phase transition, one has to solve Hamiltonian \ref{transverse_tcc} and look at the spectrum of the model.
Unfortunately, the perturbed TCC is no longer exactly solvable. However, the $X_{\fp}$ plaquette operators commute with the magnetic field and 
therefore with the full Hamiltonian Eq.~\ref{transverse_tcc}. As a consequence, the eigenvalues $\pm 1$ of all $X_{\fp}$ plaquette operators are conserved quantities and the system is block-diagonal in these quantum numbers. The $X$-type excitations are therefore frozen with a constant energy gap $2J$. It is therefore very reasonable that the relevant low-energy physics takes place in the sub-block where no $X$-type particles are present, i.e. we focus on the sub-block where all eigenvalues of $X_{\fp}$ plaquette operators are $+1$. The Hamiltonian in this sub-block is then given by 
\be \label{transverse_tcc_le}
H=-JN_{\fp}-J\sum_{\fp} Z_{\fp} -h_x \sum_{\fv} \sigma_{\fv}^x \quad .
\ee
Hence, the magnetic field in the $x$-direction only affects $Z$-type particles. 
As a consequence, the mutual fractional statistics between $X$-type and $Z$-type particles plays no role in this sub-block. 
This fact is the basis for the mapping of the perturbed TCC in this sub-block to the Baxter-Wu model in a transverse field as discussed in the next subsection. 

\subsection{Mapping of the Hamiltonian}
\label{mapping}

The elementary idea behind the mapping to the Baxter-Wu model in a transverse field is to swich from the original spin operators 
to an effective language where the eigenvalues $\pm 1$ of $Z_{\fp}$ plaquette operators are the basic degrees of freedom. 
Let us stress again that the $X_{\fp}$ operators commute with the magnetic field in the $x$-direction and the low-energy physics therefore takes place 
in the sub-block where all $X_{\fp}=1$. As a consequence, the energy of all states only depends on the eignevalues of the $Z_{\fp}$ operators. 
This fact can be formally seen by choosing the $\ket{0}^{\otimes |\fV|}$ in Eq.~\ref{tcc_ground_state} as a fully-polarized state in the $x$-direction for which 
$\frac{1}{2}(1+X_{\fp})=1$ holds trivially. As mentionned above, the spectrum of the TCC is equidistant and it depends only on the eigenvalues of the $Z_{\fp}$ 
plaquette operators in a very simple fashion. In fact, one can introduce a pseudo-spin 1/2 Pauli matrix $\tau^{z}$ where the entries on the 
diagonal exactly correspond to the eigenvalues $\pm 1$ of the $Z_{\fp}$ plaquette operators. The spectrum of the TCC is then given exactly by
\be \label{tcc_dual}
H^{\rm dual}_{\rm TCC}=-JN_{\fp}-J\sum_i \tau^{z}_i \quad ,
\ee
where $i$ refers to the center of the plaquettes $\fp$. Let us note that the centers of the plaquettes build an effective triangular lattice (see Fig.~\ref{field-pertubation}). 
We stress that this mapping keeps track of energetic properties. Any information on degeneracies is clearly lost. Indeed, the TCC is mapped
 to an effective magnetic field which has a unique polarized ground state in contrast to the topological order present for the TCC.

Next we reformulate the magnetic field term $-h_x \sum_{\fv} \sigma_{\fv}^x$ in terms of pseudo-spin variables. We therefore have to determine
 how the eigenvalues of $Z_{\fp}$ plaquette operators are affected by the action of a $\sigma_{\fv}^x$ on a certain site $\fv$. In fact, the 
operator $\sigma_{\fv}^x$ just flips the three eigenvalues of $Z_{\fp}$ plaquette operators which are attached to the site $\fv$. This is a direct consequence of the anti-commutation relation between $\sigma_{\fv}^x$ and $Z_{\fp}$ if $\fv$ is part of the hexagon $\fp$. In terms of pseudo-spin operators the flipping of three eigenvalues corresponds to $\tau^{x}_1 \tau^{x}_2 \tau^{x}_3 $ where the sites 1,2, and 3 build an elementary triangle on the effective triangular lattice.  Henceforth, the Hamiltonian \ref{transverse_tcc_le} 
is mapped to the so-called Baxter-Wu model in a transverse field:  
\be \label{ising3body}
H=-JN_{\fp}-J\sum_i \tau_{i}^z- h_x \sum_{\langle ijk \rangle} \tau_{i}{^x}\tau_{j}{^x}\tau_{k}{^x} \quad ,
\ee
where the first sum runs over the sites of a triangular lattice and the second sum runs over all triangles with sites $i$, $j$, and $k$.
One should note that the number of sites on the effective triangular lattice is equal to the number of plaquettes $N_{\fp}$ on the original honeycomb lattice. 
The magnetic field is therefore mapped to an effective three-spin interaction. Let us note that in the dual picture, 
the elementary excitations of the topological color code ($h_x=0$) correspond to simple spin flips which are static and non-interacting hardcore bosons. 

\section{Exact diagonalization}
\label{ed}
Our main goal of this paper is to understand and detect the topological phase transition between the topologically-ordered phase at low fields and 
the polarized phase present at large fields. As a first step to this end, we exactly diagonalized the Hamiltonian (\ref{ising3body}) for triangular lattices
with $N=4$, $9$ and $16$ sites using the Lanczos algorithm.
The numerical diagonalization is performed for $M\times M$ rectangular clusters where $M$ is the number of sites at each side of the rectangles and the periodic boundary conditions
are imposed at the boundaries of the clusters. 
We diagonalized the Baxter-Wu Hamiltonian in the $z$-Pauli spin basis and used the bit representation for different configurations of the spin states in the $2^N\times2^N$ Hilbert space.
In order to preserve the flow of the paper for the reader, we postpone any discussion on the results until Sec.~\ref{phase_transition}
where we use the exact diagonalization (ED) results as a crosscheck to the series expansion data to firmly support our findings from the perturbative continuous unitary transformation method.

\section{Classical approach}
\label{ca}

Following our goal, we applied a classical approach on the Baxter-Wu model in a transverse field.
Indeed, this model is much better suited for a classical calculation than the TCC in a field. The reason is that the topological phase has no classical analog. 
This is different for Hamiltonian \ref{ising3body}. The TCC is mapped to an effective field term which has a polarized ground state being well described by a classical 
configuration where all pseudo-spins point in the $z$-direction. Additionally, the ground state of the effective three-spin interactions originating from the magnetic 
field term have also a well-defined classical counterpart. Indeed, there are exactly four classical ground states for $J=0$ which are uniquely defined by a three-site 
triangular unit cell: 
$|\rightarrow\rightarrow\rightarrow\rangle$, $|\leftarrow\leftarrow\rightarrow\rangle$, $|\leftarrow\rightarrow\leftarrow\rangle$, and $|\rightarrow\leftarrow\leftarrow\rangle$
where the $\rightarrow$ ($\leftarrow$) corresponds to classical vectors poniting to the $+x$ ($-x$) direction.
It is important to realize that the effect of the effective field term in the classical limit on the four classical ground states of the Baxter-Wu model is very simple. 
The field term simply rotates every classical spin vector in essentially the same fashion, from the $x$-axis in direction to the $z$-axis in the
Cartesian coordinate system.
As a consequence, it is sufficient to use a single-site unit cell to study the Baxter-Wu model in a transverse field in the classical limit.      

Technically, we replace the vector of pseudo-spin Pauli matrices $(\tau^x,\tau^y,\tau^z)$ by a classical vectors $(\sin \theta \cos \phi, \sin \theta \sin \phi, \cos \theta)$ 
with unit length. Consequently, the classical energy of the Hamiltonian \ref{ising3body} reads as:

\be \label{classical_TCC_ising}
 \frac{E_{\rm cl}}{N_{\fp}}=-J(1+\cos \theta) -2h_x (\sin \theta  \cos \phi )^3 \quad ,
\ee
where $N_{\fp}$ is the number of sites on the triangular lattice. Note that only two angles $\phi$ and $\theta$ appear 
in this expression because a single-site unit cell has been used. In order to get the minimum of the classical energy, one finds readily $\phi=0$. 
The remaining expression only depends on $\theta$. The minimum is given either by $\theta_{\rm tcc}=0$ or by $\theta_{\rm pol}$ depending explicitly on the ratio $J/h_x$. 
The first solution $\theta_{\rm tcc}$ corresponds to a fully polarized state in $z$-direction, i.e. it refers to the polarized phase in the dual 
language or the topological phase in the original language of the perturbed TCC. This solution is definitely favored in the low-field 
limit $h_x\ll J$. In contrast, the second solution $\theta_{\rm pol}$ signals the presence of the symmetry-broken phase in the perturbed 
Baxter-Wu model or the polarized phase in the initial problem. It is certainly realized in the high-field limit $h_x\gg J$. 
The phase transition between these two solutions happens at $h_x/J=2/\sqrt{27}\approx 0.385$. The classical energy displays a 
kink and the transition is therefore first order.

The first-order nature of the phase transition in the Baxter-Wu model in a transverse field in the classical limit is already a 
hint that the topological phase transition of the TCC in a magnetic field is different compared to the toric code in a field. 
The low-energy physics of the toric code in a single field corresponds to the transverse field Ising model on the square lattice. 
The latter shows a second-order phase transition in the classical limit as well as for the full quantum problem. 
In the next section we therefore turn to the original problem of the TCC in a magnetic field and we show by high-order 
series expansion techniques and by exact diagonalizations that the quantum phase transition between the topological phase and the polarized phase 
is indeed a first-order phase transition as suggested by the classical approach.

\section{Perturbative Continuous Unitary Transformation}
\label{pcut}

The method of continuous unitary transformations (CUTs) invented by 
Wegner\cite{wegner, glazek1, glazek2} is one of the powerful techniques for diagonalizing or rather block-diagonalizing a Hamiltonian. 
Here we use the perturbative realization of the method (pCUTs) (Refs.~\cite{pcut3,pcut4}) to study the TCC in an external magnetic field. 
The central idea of CUTs is to choose a suitable generator for the unitary transformation  
and to (block-) diagonalize the initial Hamiltonian by applying an infinite sequence of successive unitary operators of particular form. 
The transformation of the Hamiltonian takes therefore place in a continuous fashion:
\be \label{flow_hamiltonian}
H(\ell)=U^{\dagger}(\ell)HU(\ell) \quad ,
\ee 
where $\ell$ is the flow parameter such that $H=H(\ell=0)$ and $H_{\rm eff}=H(\ell=\infty)$ is the effective
block-diagonal Hamiltonian. 

Considering a Hamiltonian of the form $H=H_0+xV$ where $H_0$ is diagonal and $V$ is the perturbation,
the pCUT uses the quasiparticle generator \cite{pcut3} and it maps the Hamiltonian to an 
effective one which conserves the number of quasiparticles. The pCUT is applicable to models 
where the unperturbed part of the Hamiltonian, $H_0$, has an equidistant spectrum. Additionally, the perturbation must be written in the form 
$V =\sum_{n=-N}^N T_n$ where $T_n$ increments (or decrements, if $n < 0$) the number of quasiparticles (excitations) by $n$. 
Hence, the full Hamiltonian reads:
\be
H=E_0+Q+x\sum_{n=-N}^N T_n 
\ee
where $E_0$ is the unperturbed ground state energy, $x$ is the perturbation parameter and Q is the hermitian operator counting the number of quasiparticles. 
Most importantly, 
the effective Hamiltonian $H_{\rm eff}$ conserves the number of QP's: $\comutator{H_{\rm eff}}{Q}=0$. In the following we demonstrate 
how the low-field and the high-field limit can be tackled by pCUTs. 

\subsection{Low-field limit ($h_x\ll J$)}
\label{sfl}
 
As previously outlined, we aim at a quasiparticle description for the low-energy physics of the low-field phase. 
Consequently, one would have to set up a high-order series expansion inside the topologically-ordered phase in an 
analogous fashion as it has been done for the toric code in a magnetic field \cite{vidal09,vidal09b,dusuel11}. But this
 is not necessary in the present case because one can benefit from the mapping of the perturbed TCC to the Baxter-Wu model
 in a transverse field. Indeed, the topological low-field phase corresponds to a conventional polarized phase 
in the dual picture. In the following we can therefore apply the pCUT method to Eq.~\ref{ising3body} in order to study 
 the low-field limit of the TCC in a magnetic field. 
 
The first term in Eq.~\ref{ising3body} is an effective field term having an equidistant spectrum which is considered as $H_0$ for the implementation of the pCUT. 
Hence single spin flips are the elementary 
excitations above a fully polarized ground state. The three-body Ising perturbation changes the number of spin flips by $n=\{\pm 1,\pm 3\}$. 
Therefore, Hamiltonian \ref{ising3body} can be written as:  
\be \label{qp_tcc_small}
\frac{H}{2J}=-N_{\fp}+Q+\frac{h_x}{2J} \left( T_{-3}+T_{-1}+T_{+1}+T_{+3} \right)
\ee
where $Q=\sum_i b_{i}^{\dagger} b_{i}^{\phantom{\dagger}}$ counts the number of spin flips and the $T_n$ operators are defined as
\be \label{tn_small}
\begin{gathered}
T_{+3}=\sum_{\langle ijk \rangle} b_{i}^{\dagger} b_{j}^{\dagger} b_{k}^{\dagger}=(T_{-3})^{\dagger}\\
T_{+1}=\sum_{\langle ijk \rangle} b_{i}^{\dagger} b_{j}^{\dagger} b^{\phantom{\dagger}}_{k}+ b_{i}^{\dagger} b^{\phantom{\dagger}}_{j} b_{k}^{\dagger}+ b^{\phantom{\dagger}}_{i} b_{j}^{\dagger} b_{k}^{\dagger}=(T_{-1})^{\dagger}\\
\end{gathered}
\ee
where $b_{i}^{\dagger}$ ($b_{i}^{\phantom{\dagger}}$) are creation (annihilation) operators of spin flips on site $i$. It is useful to 
display the explicit form of these bosonic operators in terms of spin operators 
\be
\tau_{i}^{-}=b_{i}^{\dagger},\ \ \tau_{i}^{+}=b_{i},\ \ \tau_{i}^{z}=1-2 n_i
\ee
where $\tau_{i}^{\pm}=(\tau_{i}^{x} \pm i\tau_{i}^{y})$.

The pCUT method maps Eq.~\ref{qp_tcc_small} to an effective Hamiltonian conserving the number of quasi particles. 
The ground state of the system corresponds to the state with zero QP ($Q=0$) and the elementary excitation is contained in the 1QP sector ($Q=1$). 
Consequently, we can compute the ground-state energy per site $\epsilon^{\rm lf}_0$, the one-particle dispersion $\omega^{\rm lf}(\vec{k})$ and the one-particle 
gap $\Delta^{\rm lf}\equiv\omega^{\rm lf}(\vec{k}=0)$. For $J=1/2$, we have calculated $\epsilon^{\rm lf}_0$ and $\Delta^{\rm lf}$ up to order 10 in the expansion parameter $h_x$
\bea \label{gs-small}
\epsilon^{\rm lf}_{0} &=& -\frac{1}{4}-\frac{1}{3} h_{x}^2-\frac{19}{54} h_{x}^4-\frac{21436}{8505} h_{x}^6- \nonumber\\
&& \frac{500690327}{21432600} h_{x}^8-\frac{74305313819}{281302875}h_{x}^{10}\\
\nonumber\\
\Delta^{\rm lf} &=& 1-12 h_{x}^2+32 h_{x}^4-\frac{134356}{81} h_{x}^6+ \nonumber\\
&& \frac{18694889252}{893025} h_{x}^8-\frac{29786981411535707}{40507614000} h_{x}^{10}\label{gap-small} \quad .
\nonumber\\
\eea

\begin{figure}
\centerline{\includegraphics[width=\columnwidth]{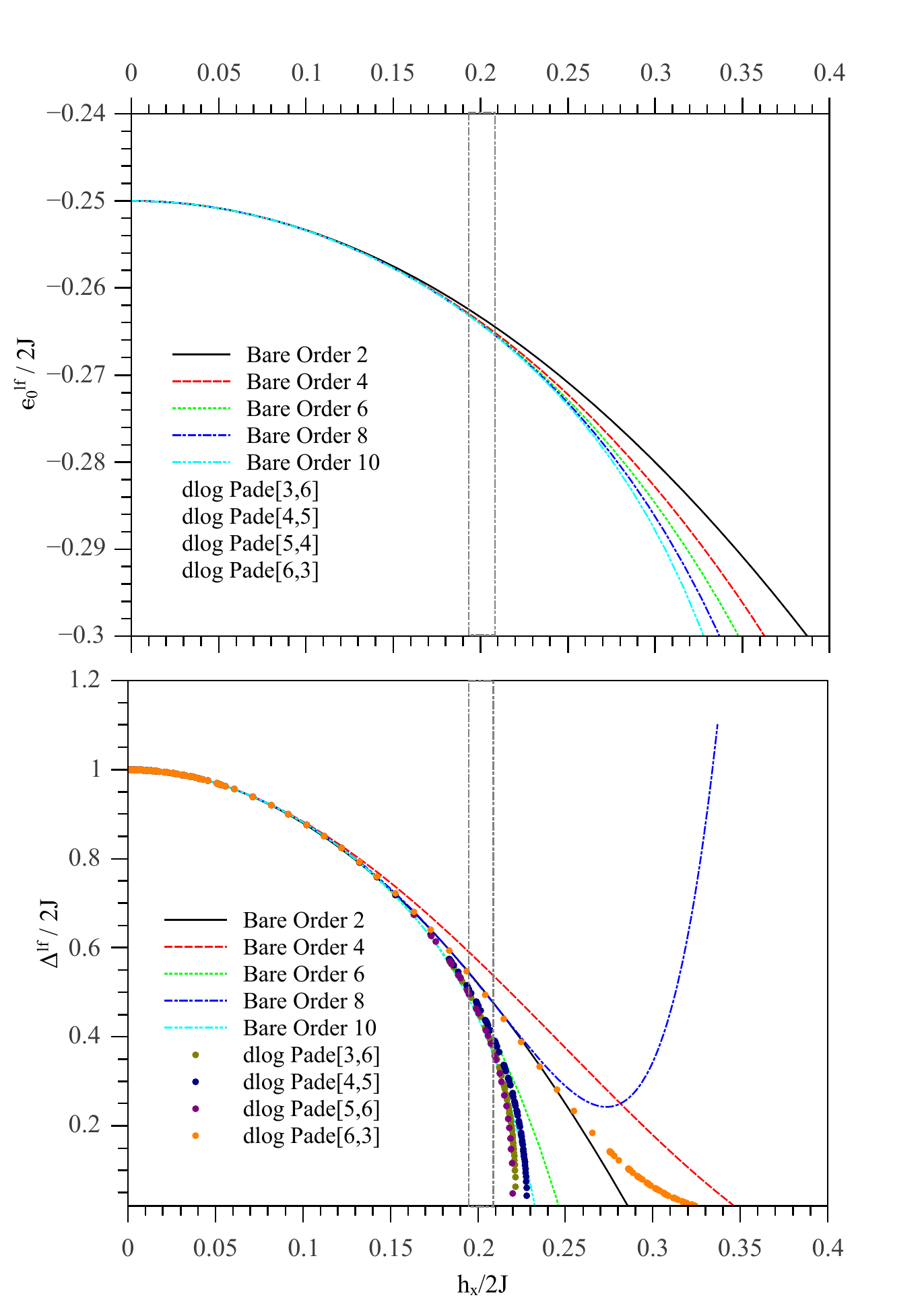}}
\caption{(Color online) Ground state energy per site $\epsilon_0^{\rm lf}$ (upper panel) and the single-particle gap $\Delta^{\rm lf}$ (lower panel) as a function of $h_x/2J$  in the low-field limit. Solid lines (symbols) represent bare series (Pad\'{e} and dlog Pad\'{e} approximants). The coupling range between the dashed vertical lines corresponds to the location of the first-order phase transition deduced from the crossing points of the ground-state energy given in Tab.~\ref{tab2}.}
\label{lowfield_extrapolation}
\end{figure}

Using standard Pad\'{e} and dlog Pad\'{e} approximants as well as the bare series of different maximal orders, the ground-state energy per site $\epsilon^{\rm lf}_{0}$ 
and the one-particle gap $\Delta^{\rm lf}$ are shown in Fig.~\ref{lowfield_extrapolation}. The bare and extrapolated series for the ground-state energy are well 
converged for $h_x/2J < 0.25$. The behavior of the one-particle gap is more complicated. The bare series of the gap has alternating signs. 
So only the bare series having a negative prefactor for the maximal order (orders 2, 6, and 10) display a reasonable behavior 
for the one-particle gap (see lower panel of Fig.~\ref{lowfield_extrapolation}). In order to get a clear picture about the nature of the phase transition from the behavior of the gap,
one need to take a look at the overall behavior of the gap in either of the low and high-field limits. Hence, we postpone any conclusion about the nature of the phase transition
to the section \ref{phase_transition} after calculating the high-field results.

\subsection{High-field limit ($h_x\gg J$)}
\label{lfl}

\begin{figure}
\centerline{\includegraphics[width=\columnwidth]{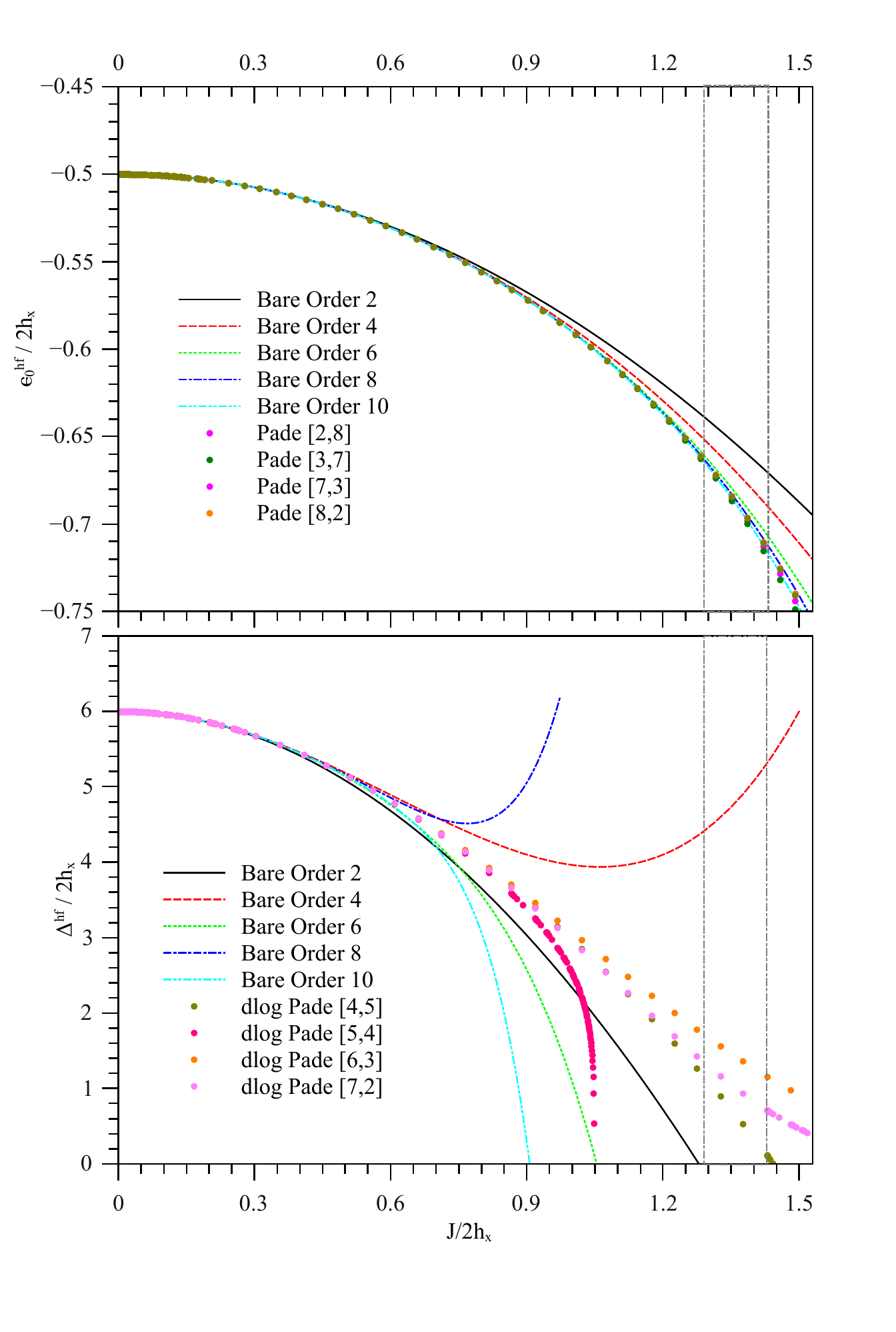}}
\caption{(Color online) Ground state energy per site $\epsilon_0^{\rm hf}$ (upper panel) and the six-particle gap $\Delta^{\rm hf}$ (lower panel) as a function of $J/2h_x$ in the high-field limit. Solid lines (symbols) represent bare series (Pad\'{e} and dlog Pad\'{e} approximants). The coupling range between the dashed vertical lines corresponds to the location of the first-order phase transition deduced from the crossing points of the ground-state energy given in Tab.~\ref{tab2}.}
\label{highfield_extrapolation}
\end{figure}

In a large enough magnetic field, the ground state of the system is in the polarized phase
along the field direction. To investigate the low-energy excitations of the system, the field term of Hamiltonian \ref{transverse_tcc} is the diagonal part, $H_0$, 
and the topological color code $H_{\rm TCC}$ represents the perturbation. Again, the overall complexity reduces strongly due to the fact that the eigenvalues of the $X_{\fp}$ plaquette 
operators are conserved quantities. The low-energy physics is expected (as already detailed above) in the sector where all these eigenvalues are $+1$. 
In this sub-block, the topological color code in a single parallel field reduces to Eq.~\ref{transverse_tcc_le}. Similarly to the dual formulation of the 
low-field limit using the Baxter-Wu model in a transverse field, the unperturbed part of the high-field limit has a fully polarized ground state and its 
elementary excitations correspond to spin-flip excitations. The perturbation flips all six spins of a hexagon. Regardless of how many particles already exist on 
the sites of a hexagon, the perturbation $Z_{\fp}$ can therefore change the number of spin flips (QP's) by $n=(0, \pm 2, \pm 4, \pm 6)$. However, 
one has to keep in mind that the $X_{\fp}$ stabilizing condition is not violated. In other words, only those processes are allowed that flip an 'even' number of 
spins on a plaquette, every time the perturbation $Z_{\fp}$ acts. As a results, processes with $n=(0, \pm 4)$ which involve an odd number of spin flips on 
plaquettes are not allowed. Finally, we can write Hamiltonian \ref{transverse_tcc_le} as follows:
\be \label{qp_tcc_large}
\frac{H}{2h_x}=-N+Q+\frac{J}{2h_x} \left( T_{-6}+T_{-2}+T_{+2}+T_{+6} \right) \quad .
\ee
The operator $Q=\sum_i b_{i}^{\dagger} b_{i}^{\phantom{\dagger}}$ counts the number of flipped spins. Table \ref{tab1} summarizes all the matrix elements of the $T_n$ operators in the language of local bosonic operators which acts on a plaquette of the lattice. Each state on a plaquette is defined via the configuration of the six spins. The vacuum of the bosonic operators corresponds to ($\ket{000000}$). Spin flips are represented by $1$'s on the sites of the plaquette.

\begin{table}[!t]
 \begin{center}
 \caption{Action of the $T_n$ operators of the Hamiltonian \ref{qp_tcc_large} on the plaquette states of the lattice.}
 \label{tab1}
   \begin{ruledtabular}
   \begin{tabular}{lccccc}
    & $T_{+6}$ &   \\
    \hline
    $b_{1}^{\dagger} b_{2}^{\dagger} b_{3}^{\dagger}b_{4}^{\dagger} b_{5}^{\dagger} b_{6}^{\dagger}  \ket{000000}$   &$\longrightarrow$    &$\ket{111111}$ \\
    \hline
    & $T_{+2}$ &   \\
    \hline
    $b_{1}^{\dagger} b_{2}^{\dagger} b_{3}^{\dagger}b_{4}^{\dagger} b^{\phantom{\dagger}}_{5} b^{\phantom{\dagger}}_{6}  \ket{000011}$   &$\longrightarrow$    &$\ket{111100}$ \\
    $b_{1}^{\dagger} b_{2}^{\dagger} b_{3}^{\dagger} b^{\phantom{\dagger}}_{4} b_{5}^{\dagger} b^{\phantom{\dagger}}_{6}  \ket{000101}$   &$\longrightarrow$    &$\ket{111010}$ \\
    $b_{1}^{\dagger} b_{2}^{\dagger} b^{\phantom{\dagger}}_{3} b_{4}^{\dagger} b_{5}^{\dagger} b^{\phantom{\dagger}}_{6}  \ket{001001}$   &$\longrightarrow$    &$\ket{110110}$ \\
    $b_{1}^{\dagger} b^{\phantom{\dagger}}_{2} b_{3}^{\dagger} b_{4}^{\dagger} b_{5}^{\dagger} b^{\phantom{\dagger}}_{6}  \ket{010001}$   &$\longrightarrow$    &$\ket{101110}$ \\
    $b^{\phantom{\dagger}}_{1} b_{2}^{\dagger} b_{3}^{\dagger} b_{4}^{\dagger} b_{5}^{\dagger} b^{\phantom{\dagger}}_{6}  \ket{100001}$   &$\longrightarrow$    &$\ket{011110}$ \\
    $b_{1}^{\dagger} b_{2}^{\dagger} b_{3}^{\dagger} b^{\phantom{\dagger}}_{4} b^{\phantom{\dagger}}_{5} b_{6}^{\dagger}  \ket{000110}$   &$\longrightarrow$    &$\ket{111001}$ \\
    $b_{1}^{\dagger} b_{2}^{\dagger} b^{\phantom{\dagger}}_{3} b_{4}^{\dagger} b^{\phantom{\dagger}}_{5} b_{6}^{\dagger}  \ket{001010}$   &$\longrightarrow$    &$\ket{110101}$ \\
    $b_{1}^{\dagger} b^{\phantom{\dagger}}_{2} b_{3}^{\dagger} b_{4}^{\dagger} b^{\phantom{\dagger}}_{5} b_{6}^{\dagger}  \ket{010010}$   &$\longrightarrow$    &$\ket{101101}$ \\
    $b^{\phantom{\dagger}}_{1} b_{2}^{\dagger} b_{3}^{\dagger} b_{4}^{\dagger} b^{\phantom{\dagger}}_{5} b_{6}^{\dagger}  \ket{100010}$   &$\longrightarrow$    &$\ket{011101}$ \\
    $b_{1}^{\dagger} b_{2}^{\dagger} b^{\phantom{\dagger}}_{3} b^{\phantom{\dagger}}_{4} b_{5}^{\dagger} b_{6}^{\dagger}  \ket{001100}$   &$\longrightarrow$    &$\ket{110011}$ \\
    $b_{1}^{\dagger} b^{\phantom{\dagger}}_{2} b_{3}^{\dagger} b^{\phantom{\dagger}}_{4} b_{5}^{\dagger} b_{6}^{\dagger}  \ket{010100}$   &$\longrightarrow$    &$\ket{101011}$ \\
    $b^{\phantom{\dagger}}_{1} b_{2}^{\dagger} b_{3}^{\dagger} b^{\phantom{\dagger}}_{4} b_{5}^{\dagger} b_{6}^{\dagger}  \ket{100100}$   &$\longrightarrow$    &$\ket{011011}$ \\
    $b_{1}^{\dagger} b^{\phantom{\dagger}}_{2} b^{\phantom{\dagger}}_{3} b_{4}^{\dagger} b_{5}^{\dagger} b_{6}^{\dagger}  \ket{011000}$   &$\longrightarrow$    &$\ket{100111}$ \\
    $b^{\phantom{\dagger}}_{1} b_{2}^{\dagger} b^{\phantom{\dagger}}_{3} b_{4}^{\dagger} b_{5}^{\dagger} b_{6}^{\dagger}  \ket{101000}$   &$\longrightarrow$    &$\ket{010111}$ \\
    $b^{\phantom{\dagger}}_{1} b^{\phantom{\dagger}}_{2} b_{3}^{\dagger} b_{4}^{\dagger} b_{5}^{\dagger} b_{6}^{\dagger}  \ket{110000}$   &$\longrightarrow$    &$\ket{001111}$ \\
    \hline
    & $T_{-2}=(T_{+2})^{\dagger}$ &   \\
    \hline
    & $T_{-6}=(T_{+6})^{\dagger}$ &   \\

    \end{tabular}
  \end{ruledtabular}
 \end{center}
\end{table}

Bearing the stabilizing condition of $X_{\fp}$ in mind, we would like to emphasize that 
although the single spin flip channel is the lowest excited state for the diagonal 
part of Hamiltonian \ref{qp_tcc_large}, it is not the relevant one for the phase transition happening between the topological phase and the polarized phase. 
Keeping the $X_{\fp}$ operators in their ground state enforces that the true lowest excitation of
the system is the 6-QP sector with an energy gap $6$ having all six particles located on one hexagon. The next excited state is created 
by acting on the neighboring plaquette with an energy gap $8$, third one with $10$ and so forth. 
Setting $h_x=1/2$, the ground state energy per site and the six-particle gap $\Delta^{\rm hf}\equiv\omega^{\rm hf}(\vec{k}=0)$ of the TCC in the large-field limit read: 

\bea \label{gs-large}
\epsilon^{\rm hf}_{0} &=& -\frac{1}{2}-\frac{1}{12}J^2-\frac{1}{216}J^4-\frac{19 }{9720}J^6-\frac{1133 }{3732480}J^8 - \nonumber\\
&& \frac{12026279}{105815808000}J^{10}\\
\nonumber\\
\Delta^{\rm hf} &=& 6-\frac{11}{3}J^2+\frac{44}{27} J^4-\frac{413}{144} J^6+\frac{20157041}{3499200} J^8- \nonumber\\
&& \frac{1446718370831}{105815808000} J^{10}\label{gap-large} \quad .
\nonumber\\
\eea

The structure of the $T_n$ operators implies that the six QPs have to hop around together on the lattice,
from one plaquette to the other, and can not disperse individually. This situation is similar to the low-field limit
where particles are created on plaquettes and hop around on an effective triangular lattice. In the high-field limit one can 
think of the six spin flip quasiparticles as a single compound object which emerges on the plaquettes and their kinetics takes place again 
on a triangular lattice. 

Figure \ref{highfield_extrapolation} illustrates the ground-state energy per site $\epsilon^{\rm hf}_{0}$ and the six-particle gap $\Delta^{\rm hf}$ as a function of 
$J/2h_x$. The bare and extrapolated ground-state energy curves are well converged for $J/2h_x <1.4$. Similar to the low-field gap, the convergence of the bare six-particle gap is very
 poor due to the alternating sign structure of the series. Again, we emphasize that by looking at the full phase diagram along 
the whole ranges of field strength, one can obtain a better understanding about the nature of the phase transition. 
In the next section we present strong evidences for a first-order phase transition.

\section{Phase transition}
\label{phase_transition}

\begin{figure}
\centerline{\includegraphics[width=\columnwidth]{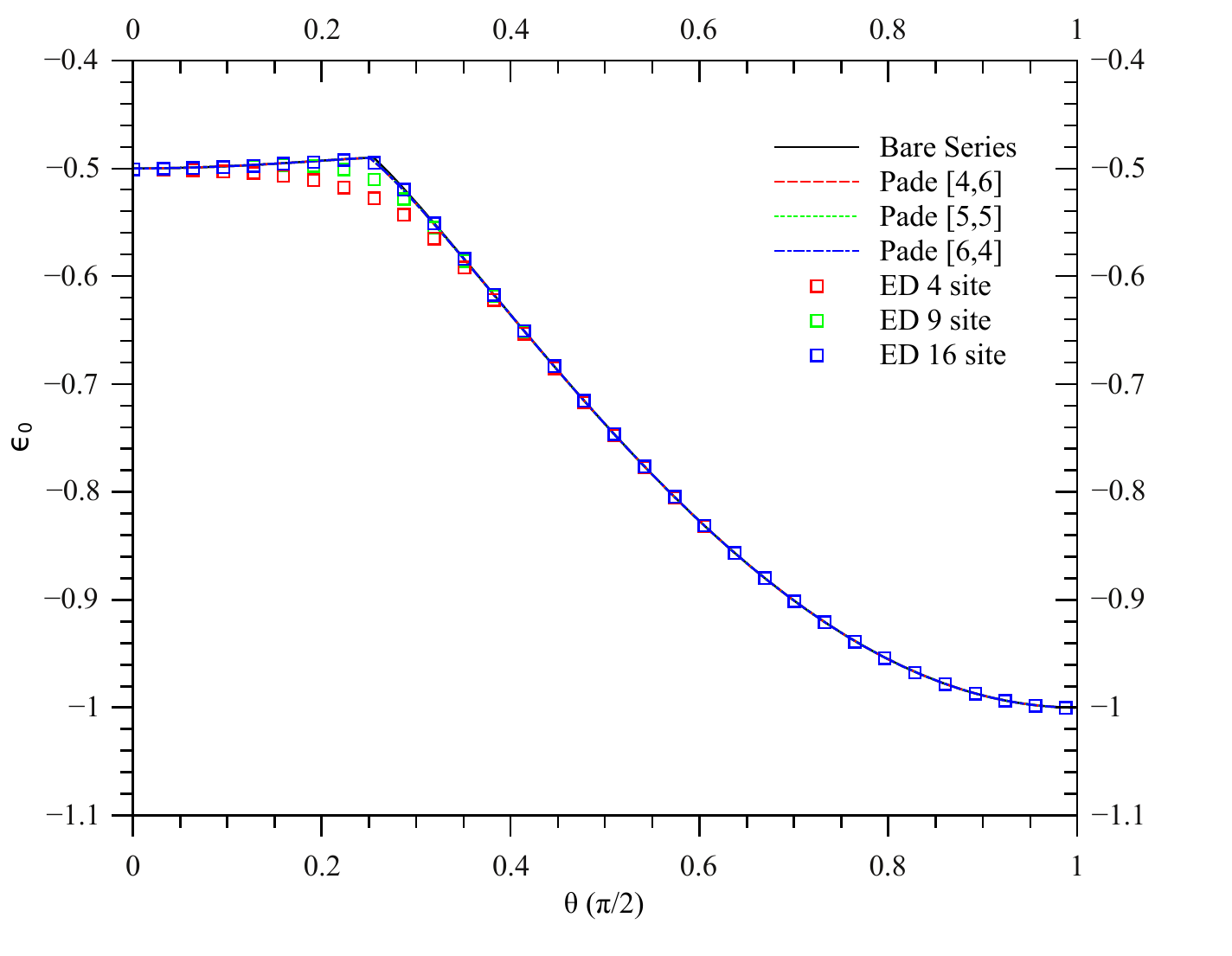}}
\caption{(Color online) The ground state energy per site $\epsilon_0$ as a function of $\theta$. Solid line corresponds to the bare series of order 10. 
Circles denote different Pad\'{e} approximants and the squares represent the exact diagonalization results for different lattice sizes.}
\label{gsps}
\end{figure}

\begin{figure}
\centerline{\includegraphics[width=\columnwidth]{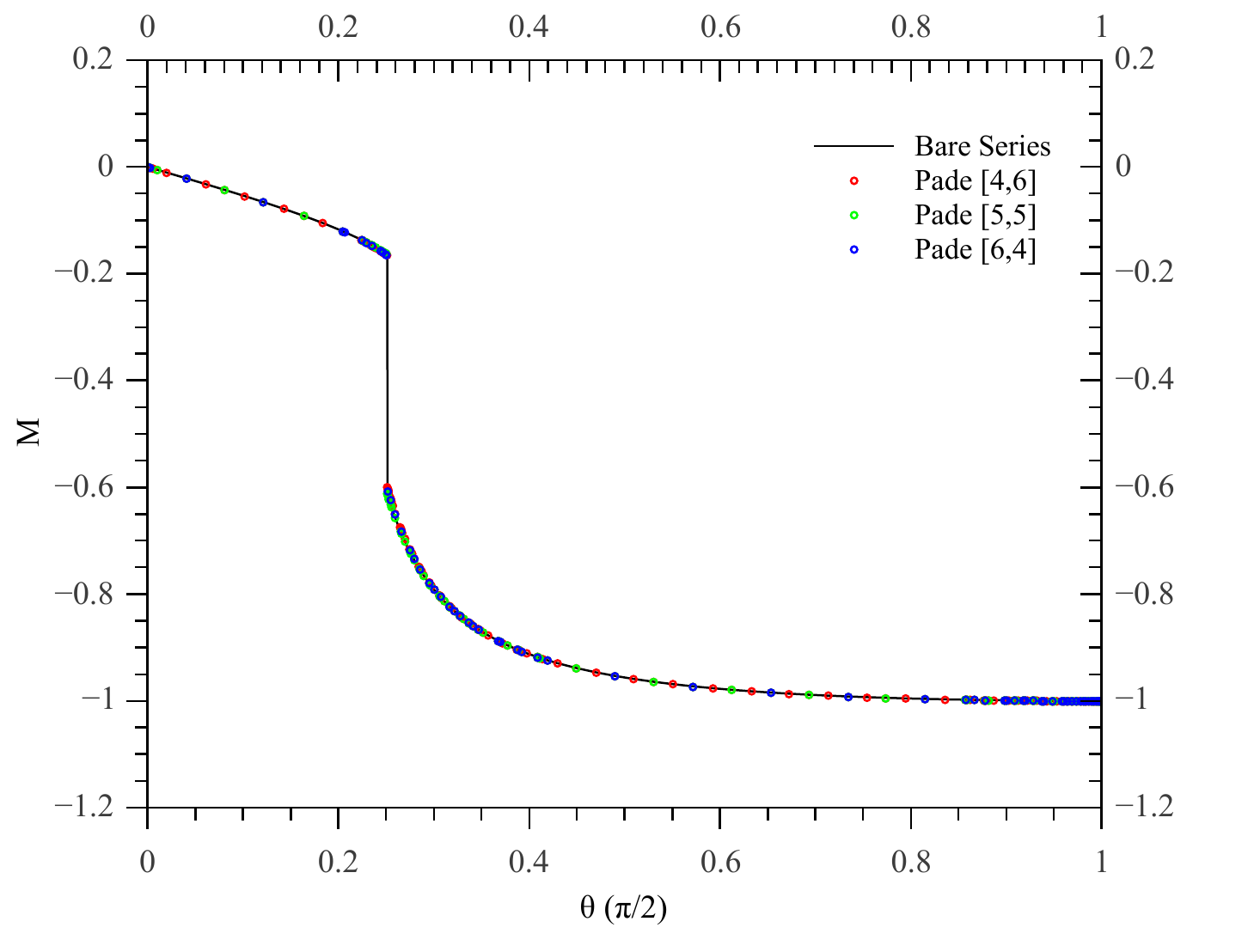}}
\caption{(Color online) The magnetization $M$ of the system as a function of $\theta$. Solid line corresponds to the bare series of order 10. 
Symbols denote different Pad\'{e} approximants.}
\label{magnetization}
\end{figure}

\begin{figure}
\centerline{\includegraphics[width=\columnwidth]{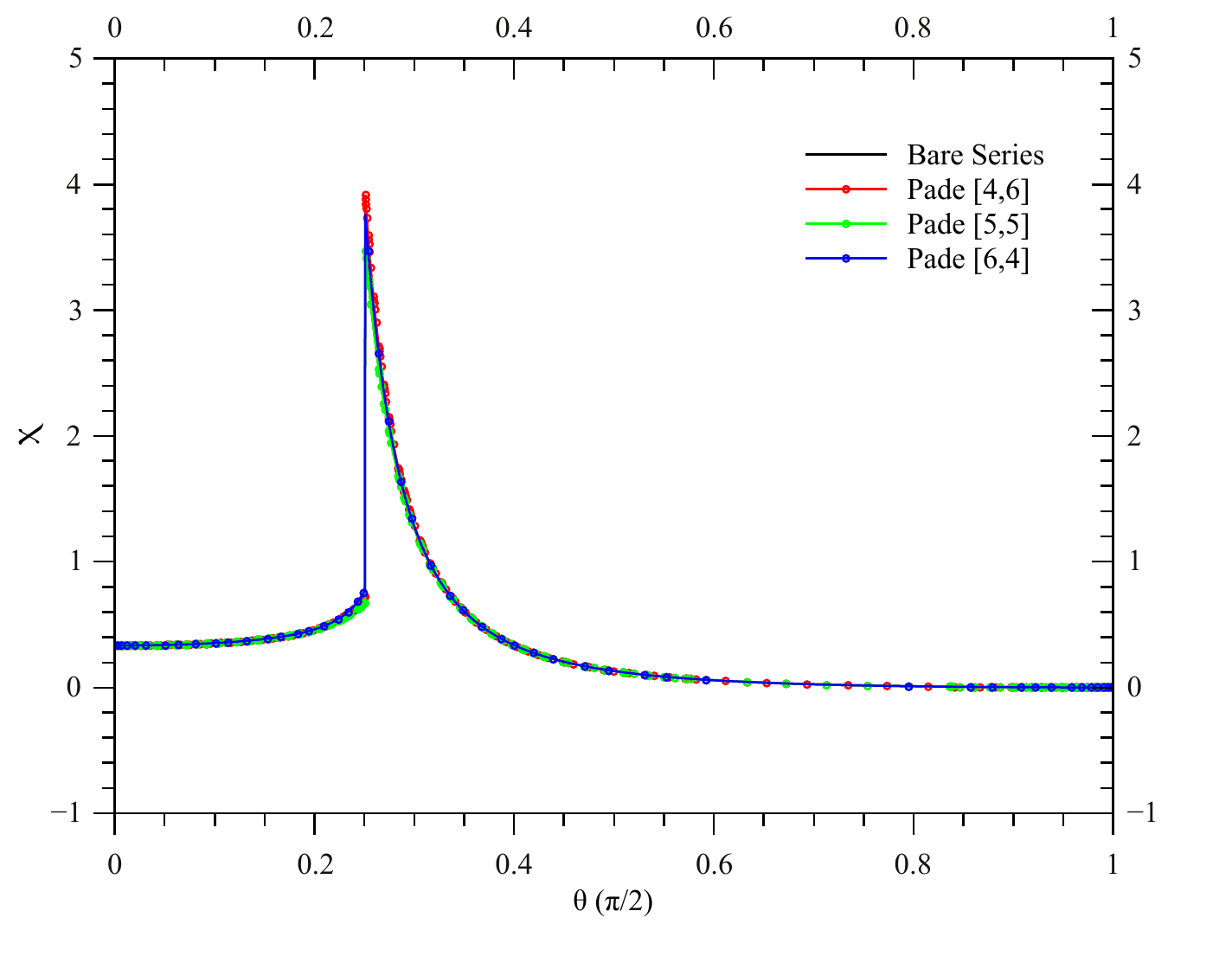}}
\caption{(Color online) The ground state susceptibility $\chi$ of the system as a function of $\theta$. Solid line corresponds to the bare series of order 10. 
Symbols denote different Pad\'{e} approximants.}
\label{susceptibility}
\end{figure}

\begin{figure}
\centerline{\includegraphics[width=\columnwidth]{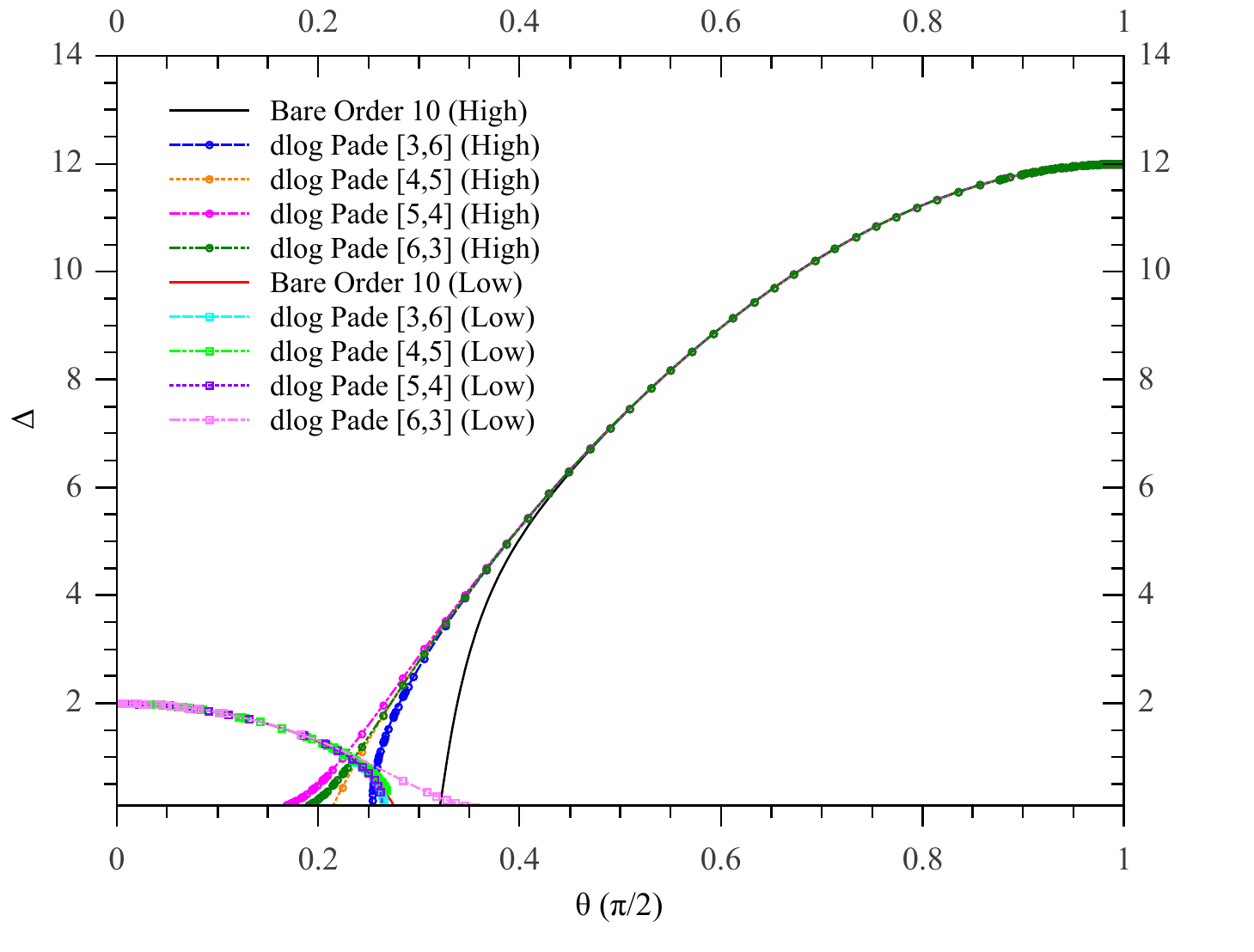}}
\caption{(Color online) Low- and high-field gap $\Delta$ as a function of $\theta$ for different dlogPad\'{e} approximants.}
\label{gap}
\end{figure}

To investigate the nature of the phase transition, we study the behavior of the ground-state energy and the elementary gap in the whole range of parameters 
by combining our results of the low- and high-field expansions. Additionally, we performed exact diagonalizations for the Baxter-Wu model in a transverse field.
Choosing $h_x=\sin \theta$ and $J=\cos \theta$, one can join low- and high-field quantities 
in the same plot to get a complete picture (Here one has to be careful not to mistake $\theta$ with the one which was defined in Sec.~\ref{ca}. The former was used as the angle which
minimizes the classical energy). 
 
We start our discussion with the ground-state energy per site which is shown in Fig.~\ref{gsps}. We find that the ground-state energy per site of the low-
 and high-field limits cross for a certain value $\theta_{\rm c}$ giving a kink in the ground-state energy per site. Most importantly, the location of this kink 
$\theta_{\rm c}\approx 0.394$ is essentially the same for various Pad\'{e} approximants (see Table.~\ref{tab2}) as well as for the bare series. The latter can be 
clearly seen from Figs.~\ref{lowfield_extrapolation} and \ref{highfield_extrapolation} due to the fact that the crossing point $\theta_c$ is located in the range of 
parameters for both limits where even the bare series of the ground-state energy is well converged. 
The analysis of the ground-state energy per site points is therefore consistent with a first-order
phase transition taking place at $\theta_{\rm c}$.
Figure \ref{gsps} is further supplemented with finite-size exact diagonalization results for different lattice sizes. 
As one can clearly see, the ED data for the 16-site cluster 
shows a kink at the same critical value as the series-expansion data, certifying that the nature of the transition is of first-order. 
The ED and pCUT results are also in full accordance with the classical approximation done for the Baxter-Wu model in a transverse field.
Surprisingly, the classical phase transition point $\theta^{cl}_{\rm c}\approx0.367$ is very close to $\theta_{\rm c}$.

We also calculated magnetization $M=d\epsilon_0/dh_x$ and the ground-state susceptibility $\chi=-d^2\epsilon_0/dh_{x}^{2}$ from the ground-state energy via the Feynman-Hellman theorem. These two quantities are shown in Figs.~\ref{magnetization} and \ref{susceptibility}. 
The first-order nature manifests itself in a jump in the magnetization at $\theta_{\rm c}$. We stress that the height of the jump is essentially the same for the
 bare series and for various Pad\'{e} approximants confirming quantitatively the location and the nature of the phase transition.

Finally, we discuss the behavior of the energy gap $\Delta$ as a function of $\theta$ which is displayed in Fig.~\ref{gap}. As already discussed above, 
 the convergence of the bare series is poor due to the alternating sign of the series in both limits. Additionally, dlogPad\'{e} extrapolation of both limits
 shows no clear tendency in favor of a gap closing signaling a second-order phase transition. In contrast, most of the extrapolants of both limits cross rather close
 to $\theta_{\rm c}$ as can be seen from Table.~\ref{tab2} which is in full agreement with a first-order phase transition.  

\section{Conclusion}
\label{conclud}

\begin{table}[t]
 \begin{center}
 \caption{Crossing points $\theta_{\rm c}$  of the low- and high-field gap $\Delta$ (left column) and the ground state energy per site $\epsilon_0$ (right column) signaling a first-order phase transition point.}
 \label{tab2}
   \begin{ruledtabular}
   \begin{tabular}{lcccc}
    $\Delta$ & $\theta_c$  & \vline & $\epsilon_0$ & $\theta_c$\\
    \hline
    
      dlog Pad\'{e} $[2,7]$   & 0.401  & \vline  &  Pad\'{e} $[2,8]$&  0.394\\
      dlog Pad\'{e} $[3,6]$   & 0.400  & \vline  &  Pad\'{e} $[3,7]$&  0.395\\
      dlog Pad\'{e} $[4,5]$   & 0.375  & \vline  &  Pad\'{e} $[4,6]$&  0.394\\
      dlog Pad\'{e} $[5,4]$   & 0.358  & \vline  &  Pad\'{e} $[5,5]$&  0.394\\
      dlog Pad\'{e} $[6,3]$   & 0.370  & \vline  &  Pad\'{e} $[6,4]$&  0.394\\
      dlog Pad\'{e} $[7,2]$   & 0.372  & \vline  &  Pad\'{e} $[7,3]$&  0.395\\
    
    \end{tabular}
  \end{ruledtabular}
 \end{center}
\end{table}

The topological color code is proposed as a platform where topological quantum computation can be done in the sub-space of the different topological ground states. 
It is therefore interesting to understand how external perturbations such as a magnetic field affect this topological phase and what kind of topological phase transitions do occur. 
In this work we studied the robustness of the topological color code in a single parallel magnetic field on the honeycomb lattice. 
Due to the fact that one sort of plaquette operators still commutes with the full Hamiltonian, one type of elementary excitations is frozen and the 
relevant low-energy physics takes place in the sub-block where only one type of elementary excitations is present.

We have shown that the perturbed color code in this sub-block is isospectral to the Baxter-Wu model in a transverse field on the triangular lattice. 
To the best of our knowledge this model has not been studied before in two dimensions. 
Using high-order series expansion about the limit of low and high fields, 
we find evidences that the phase transition between the topologically ordered phase and the polarized is first order and takes place at the 
critical field $h_x/J\approx0.383$. 
Our findings are further in full agreement with the exact diagonalization results of the Baxter-Wu model in a transverse field. 
The behavior of the TCC in a magnetic field is therefore different to the analogous problem of the perturbed toric code. 
The latter displays a second-order quantum phase transition at the critical point $h_x/J\approx0.328$ for the case of a single parallel field where 
the low-energy physics of this problem is given by the conventional Ising model in a transverse field on the square lattice. 
Our results therefore suggest that the topological color code is more robust than the toric code. 

\section{Acknowledgements}
KPS acknowledges ESF and EuroHorcs for funding through his EURYI. MK acknowledges financial support from ARO Grant W911NF-09-1-0527 and NSF Grant DMR- 0955778. 
SSJ also acknowledges KPS for hospitality during his stay at the TU Dortmund and Michael Kamfor for fruitful
discussions.

\end{document}